# Rotation-tuned single hexagonal air cavity assisting in third-harmonic generation via hybrid modes

Hao Song[1,2,*], Junmin Deng[1,2], Yu Chen[1,2], Yanming Sun[3], Ming-Chun Tang[4,*] and Guo Ping Wang[3]

[1]College of Physics and Electronic Information Engineering, Neijiang Normal University, Neijiang, 641100, China

[2] Neijiang Optoelectronic Devices Engineering Research Center, Neijiang, 641100, China

[3]College of Microelectronics and Communication Engineering, Chongqing University, Chongqing 400044, China

[4]College of Electronics and Information Engineering, Shenzhen University, Shenzhen, 518060, China

*Email: haosongnju@njtc.edu.cn, tangmingchun@cqu.edu.cn

**Abstract**

A fillable air cavity with a high quality ($Q$) factor and large-scale electric field confinement is highly desired in many optical applications. Yet, it remains challenging due to the dielectric transparency and metal loss in optical and near-infrared regimes. Here, we present a rotated hexagonal air cavity embedded in an Ag-air-Ag waveguide. Under near-infrared excitation, evanescent waves tunnel into the cavity. In addition to the whispering gallery mode and surface plasmon polaritons, the cavity also induces Fabry-Pérot (FP) resonance, whose orientation is tunable via cavity rotation. Thus, our cavity possesses much stronger field confinement and higher $Q$ than a circular cavity lacking FP resonance. The waveguide exhibits suppressed backward reflection filtering and Fano-type lineshapes. Then, integrating a silicon cylinder into the cavity, we demonstrate linear tuning of Mie resonances via radius adjustment. When the electric dipole (ED) resonance is excited, energy is predominantly confined within the cylinder. Different Mie modes will change the orientation of the FP resonance. Furthermore, the hybrid modes with ED resonance induce the third-harmonic wave of green light. These findings offer a promising strategy for designing high-$Q$ air cavities for next-generation multifunctional electro-optical devices.

## 1 Introduction

Optical cavities characterizing high quality ($Q$) factors and exceptional optical confinement are essential components for various applications such as lasers [1], sensors [2], and nonlinear optics [3]. Generally, the critical factor for determining $Q$ is the loss of the cavity, including the radiation and intrinsic non-radiation loss [4-7]. Therefore, researchers have dedicated themselves to achieving a high $Q$ cavity by minimizing the non-radiation and radiation loss. The high refractive index dielectric materials are promising candidates for the cavity due to their lossless properties in the near-infrared regime [8-10]. For instance, with exciting whispering gallery mode (WGM) [11-13], the electromagnetic (EM) energy is confined in the high refractive index lithium niobate (LN) microcavity with a $Q$ of about $10^6$, enabling the acousto-optic modulation [14]. Furthermore, the LN cavity with the polygon modes generates a soliton microcomb [11]. In addition, silicon microresonators have been verified to possess a $Q$ factor exceeding one billion [12, 15], making them suitable for soliton microwave oscillators [16]. The free-standing cavity has been proposed to further improve the $Q$ value by blocking the leakage channel of the EM energy from the cavity to the substrate [4, 17]. Nevertheless, these cavities put forward high requirements for fabrication technology, such as femtosecond laser micromachining, electron beam lithography (EBL), and chemomechanical polishing [17], to avoid energy loss caused by defects [18]. Inconveniently, the solid cavities without other fillable space are unsuitable for switchable and versatile optical response scenarios because their responses are fixed as long as the structures are determined.

Thus, a high $Q$ air cavity provides an opportunity to address this constraint. Previous works proved that high-index silicon or germanium plates patterned with air holes or slits can induce strong in-plane energy confinement within the air domain [6, 19-23]. Photonic bound states in the continuum (BIC) are the underlying physical mechanism of the confinement with high $Q$. BIC, as a localized bound state, occurs within the radiation continuum, which confines the EM energy perfectly in the plane and possesses an ultrahigh $Q$ by symmetry-breaking perturbations. Therefore, as mentioned before, BIC achieves a high-$Q$ cavity by suppressing radiation loss. Nevertheless, these components have impediments in applications such as integrated photonic chips due to the large size of the 2D infinite extension. Furthermore, the high $Q$ also depends on the structure symmetry, which is unsuitable for the single air cavity. Therefore, a small-sized air cavity is a promising candidate for overcoming these limits. Previous work has investigated the field confinement inside the air cavity enclosed by a silicon bulk [24]. The response wavelength relies on the depth and width of the hole, forming the Mie void modes in the ultraviolet or the visible short-wave regimes [24, 25]. However, it is difficult to realize the near-infrared confinement in an air cavity at will because the surrounding dielectric medium is almost transparent to near-infrared waves.

Next, we pay attention to the single air cavity enclosed by metallic boundaries. When exciting surface plasmon polaritons (SPPs) or localized surface plasmon (LSP), metal structures can significantly enhance the electric field in the nearby air, albeit unable to avoid energy dissipation [26-29]. The field enhancements typically emerge on nanoscale regions close to the surface, resulting in so-called 'hot spots'. Diversely, the single air cavity with metallic boundaries exciting WGM enables energy confinement in a relatively large-scale 'hot area' [30, 31]. Some studies suggested that air slits or holes on the metal slabs under free space incidence can empower extraordinary optical transmission [32], photocatalysis [33], single-molecule detection [34], etc. However, due to the metallic absorption losses, the energy confinement and $Q$ value of the single air cavity are limited when only the WGM and SPPs are excited.

In contrast, the polygon modes induced by weak perturbations [11], the circular side square resonator through geometric deformation [35], and square microcavity lasers [36, 37], can confine light inside the cavities with much higher $Q$ values, which originates from the less scattering loss and reducing defect influence from the cavity sidewall [38, 39]. Additionally, the angular perturbation method has been proven to improve the $Q$ factor and the robustness of the photonic crystal slab with air slit arrays [23]. Inspired by these findings, to address the limits mentioned above, in this work, we proposed a rotated hexagonal air cavity bounded by silver walls, which can excite hybrid resonant modes.

Specifically, the cavity is integrated into an Ag-air-Ag waveguide. With near-infrared wave grazing incidence, the waveguide induces SPPs. Simultaneously, the evanescent electric field tunnels into the air cavity and then is confined on a large scale when WGM emerges. Beyond these modes, the hexagonal cavity also excites significant FP resonance between one set of opposite sidewalls. Interestingly, the orientation of the FP resonance is switchable as the cavity rotates. Compared to the conventional circular cavity, the hexagonal cavity with the hybrid modes can enhance the confinement intensity and the $Q$ value by about 1.69 and 1.74 times, respectively. Moreover, the hybrid modes lead to typical Fano response curves and filtering with less backward reflection. Then, our cavity possessing a large-scale

'hot area' can assist in exciting various Mie resonances of a silicon cylinder embedded into the cavity. In particular, the maximum energy confinement occurs inside the cylinder at ED resonance or inside the remaining air at EQ resonance, with resonant wavelengths exhibiting linear dependence on cylinder radius. Subsequently, highly efficient third-harmonic generation (THG) from the cylinder is achieved at the ED resonance. Finally, we consider the impacts of the refractive index of the cylinder. Our findings facilitate the design of a single air cavity with strong field confinement and high $Q$, highlight a new strategy for manipulating optical responses through the hybrid modes, and provide an accessible component with promising applications in integrated chip optical communication, spectrometer, laser, etc.

## 2 Results and discussions
### 2.1 Rotated hexagonal air cavity

We start by designing a circular air cavity integrated into an Ag-air-Ag waveguide, as shown in Figure 1(a). The white domains are air with a unity refractive index, and the gray is Ag. The permittivity of Ag is satisfied with the Drude Model expressed by [40, 41]

$$\varepsilon_m = \varepsilon_\infty - \frac{\omega_p^2}{\omega^2 + i\omega\gamma}, \tag{1}$$

$\varepsilon_m$ being the permittivity of 3.7 at infinity, $\omega_P$ being the plasmon oscillation frequency of $1.38 \times 10^{16}$ rad/s, $\gamma = 2.37 \times 10^{13}$ rad/s being the damping oscillation frequency, and incident angular frequency $\omega$. $r_0$, $w$, and $g$ are the radius of the circular boundary, the width of the straight-line air waveguide, and the minimal gap between the circular cavity and the line waveguide, respectively. The EM wave is incident from the left port of the line waveguide (labeled by '*in*') and is output at the right port ('*out*').

For efficiently exciting the SPPs, the incident TM wave propagates along the *x*-direction and is polarized in the *y*-direction. As proof of SPPs generation, we assume an interface at *y*=0 between the semi-infinite Ag substrate and the straight-line air waveguide. Owing to the silver's negative permittivity in the visible and near-infrared regimes [41, 42], total internal reflection is achieved at the interface under grazing incidence. According to the Maxwell Equations and the boundary condition, the electric field in the air (subscript 1) and in the metal (subscript 2) can be written as

$$E_{y1(2)} = C_{1(2)} e^{i\beta x} e^{-\sqrt{\beta^2 - k_0^2 \varepsilon_{1(2)}} |y|}. \tag{2}$$

Here, $\beta$ is the propagation constant, $k_0$ is the wavenumber in vacuum, $\varepsilon_1$ and $\varepsilon_2$ are the permittivities, and $D_1$ and $D_2$ are amplitude coefficients [43]. As a result, a surface wave propagates along the *x*-direction while the evanescent field decays in the *y*-direction, i.e., the SPPs. The attenuation length can be defined as $1/\sqrt{\beta^2 - k_0^2 \varepsilon_{1(2)}}$. The $E_y$ tunnels through the metal into the circular cavity, resulting in the WGM. Therefore, the parameters $w$ and $g$ will affect the evanescent wave intensity in the cavity. For simplicity, the $w$=50 nm and $g$=15 nm are fixed throughout this work. We employ the commercial software COMSOL based on the finite element method (FEM) to perform the numerical analysis [44]. The perfectly matched layer and scattering boundaries enclose the entire structure to absorb the outward radiation. In the simulation, the three-dimensional (3D) structure is thick enough to be replaced safely by a 2D one.

We aim to realize electric field confinement in the air cavity near the optical communication wavelength of 1550 nm. To assess intuitively the confinement, a normalized parameter nUc referring to the average energy density in the cavity was proposed and written as [45, 46]

$$nUc = \frac{\int_s n^2 |\mathbf{E}|^2 \, d^2\mathbf{r}}{S} \quad , \tag{3}$$

$S$, $\mathbf{E}$, and $n$ are the cavity's area, electric field, and refractive index. $|\mathbf{E}|$ was treated as the electric field norm (nE) during the analysis. It should be highlighted that the incident electric field amplitude is 1, so the electric field in the cavity will be enhanced nE times. $\mathbf{r}$ is the position vector. Then, the nUc curve for the $r_0$ is represented in Figure 1(b). The first peak with nUc= 4.73 occurs at $r_0$=421 nm. Thus, the first (or fundamental) WGM appears at this peak. Subsequently, a higher-order WGM emerges at the second peak when $r_0$=714 nm. The inset shows the nE distribution at this peak, and the maximum nE is 2.24, which verifies the higher-order mode. Alternatively, the first and second peaks represent the dipole-like and quadrupole-like Mie void modes, respectively [24]. However, the nUc of the higher-order WGM is much smaller than the fundamental one. Therefore, we focus on the fundamental mode to achieve high confinement intensity in the cavity.

Next, Figure 1(c) illustrates the nUc for the incident wavelength ($\lambda$) at the fundamental WGM. Point $P_0$ denotes the $\lambda$=1551.4 nm at the peak of nUc=5.11. To verify the confinement further, we discuss the $Q$ factor of the nUc spectra as a function of $\lambda$. $Q = \omega_0/\Delta\omega$, $\omega_0$ is the frequency at the peak, and the $\Delta\omega$ is its full width at half maximum (FWHM) [15, 45, 47]. Here, the $Q$ is about 156.70. The reflectivity (R, green curve), transmissivity (T, blue curve), and absorption (A, red curve) spectra of the fundamental WGM as a function of $\lambda$ are depicted in Figure 1(d). The R and T are defined based on the energy flux at the input and output ports, written as [48, 49],

$$R = \int_l \frac{p_r}{p_i} dl \; , \; T = \int_l \frac{p_t}{p_i} dl \; , \tag{4}$$

$l$ denotes the port width, $p_r$ and $p_t$ are the outflow time-average energy densities from the reflected and transmitted ports, respectively,

$$p_{r,t} = \left| \frac{1}{2} \text{Re}(\mathbf{E}_{r,t} \times \mathbf{H}^*_{r,t}) \right| \; . \tag{5}$$

$p_i$ is the incident energy density expressed by

$$p_i = \frac{1}{2} c\varepsilon |\mathbf{E}|^2 \; , \tag{6}$$

$c$ is the light speed, $\varepsilon$ is the dielectric constant of the environment. In addition, A is defined as the ratio between loss and incident energy, described by

$$A = \frac{\int_{S_{tot}} U_{loss} d\mathbf{r}^2}{\int_l p_i dl} \; . \tag{7}$$

Here, $U_{loss}$ is the total power dissipation density, $S_{tot}$ is the cross-section of the total structure. Thus, a dip of T and peaks of R and A curves occur near 1550 nm. At $P_0$, A=0.44 is near the maximum value, R=0.46, and T=0.082. Thus, the filtering feature at $P_0$ originates from the pronounced reflection and absorption.

Eventually, the asymmetric lineshapes of R, A, and T indicate the generation of Fano resonance [7, 50, 51]. It is worth underscoring the famous Fano shape formula [50, 52]

$$\rho(E) \propto \frac{(q+\Omega)^2}{1+\Omega^2} \; . \tag{8}$$

The reduced energy $\Omega = 2(N-N_0)/\Gamma$, $N_0$ and $\Gamma$ are the resonant energy and width, respectively. q is the Fano parameter, which indicates the degree of asymmetry of the curve. According to the signs and size of q, it is easy to deduce the actual resonant frequency location and the approximate lineshape and vice versa [50, 53]. Therefore, the q of the T, R, and A curves are about 0, -1, and over 1, respectively.

Then, Figure 1(e) demonstrates the nE distribution at $P_0$. A significantly large-scale evanescent wave hot area appears in the air cavity, and the maximum electric field enhancement is 3.13. These results agree well with the generation of the fundamental WGM. Furthermore, remarkably sharp standing wave stripes occur in the straight-line air waveguide in the reflected direction, which originates from the constructive interference between the incident wave and the opposite wave induced by the cavity [54]. While significant energy is confined in the cavity and the line waveguide, there is no absorption due to the air being lossless. To gain deeper insight into the physical origin of the absorption, Figure 1(f) indicates the current density norm nJ at $P_0$. Consequently, the current is distributed on the metal side along the interfaces. The maximum nJ occurs in the gap between the cavity and the waveguide. According to Ohm's law, the current density is calculated as $\mathbf{J}=\sigma\mathbf{E}$. $\sigma$ is the conductivity. Thus, the nJ is positively correlated with the nE. Based on formula (2), the confined evanescent wave enhances the nE in the cavity. Consequently, the time-averaged rate of dissipation energy is expressed by [55]

$$U_{loss} = \frac{1}{2}\mathbf{J}\cdot\mathbf{E}^* = \frac{1}{2\sigma}|\mathbf{J}|^2 \quad . \tag{9}$$

Therefore, the absorption originates from the enhanced evanescent electric field in the metal, especially in the gap.

To protect components and improve efficiency, the backward reflection of on-chip optical communication should be minimized as much as possible. Nevertheless, the circular cavity generates more backward reflection than absorption and transmission. Therefore, we aim to achieve a high $Q$ cavity featuring maximum absorption with low reflection. We design an equilateral hexagonal air cavity with a side length of $l$, as presented in Figure 2(a). Except for $l$, other geometric and material parameters remain unchange. Figure 2(b) displays the nUc as a function of $l$ under $\lambda$=1550 nm. The fundamental WGM appears at $l$=458 nm, and nUc is about 7.98. Figure 2(c) demonstrates the nUc concerning $\lambda$ when $l$=458 nm. The peak of nUc=8.15 occurs at point $P_1$, which denotes $\lambda$=1549.6 nm. Thus, the $Q$ value is about 212.42, much larger than the circular cavity. The $\lambda$-dependent R, A, and T spectra of the hexagonal cavity under $l$=458 nm are shown in Figure 2(d). At $P_1$, the maximum absorption is A=0.62, while a small R=0.18. Then, Figure 2(e) draws the nE distribution at $P_1$. A noticeable electric field hot area occurs in the air cavity, which indicates the generation of the fundamental WGM. The maximum nE increases to 3.95. Moreover, the FP resonance leads to the significant electric field confinement between the cavity's top and bottom boundaries, which is proved by the distance between the two boundaries approximately equal to $\lambda/2$. Subsequently, the nJ distribution at $P_1$ is illustrated in Figure 2(f). Similarly, the SPPs propagate along the boundaries, and the evanescent electric field in metal causes remarkable absorption. In addition, the nJ of the corner is larger than the other locations. As a result, compared to the circular cavity, the hybrid modes of WGM, SPPs, and FP resonance prompt more significant electric field confinement, and the stronger evanescent electric field generates more absorption. Furthermore, hybrid modes result in more complex asymmetric Fano lineshapes of R, A, and T in Figure 2(d).

Next, we try to rotate the hexagonal cavity with $l$=458 nm. Figure 3(a) shows the schematic of the rotated cavity. Here, $\theta$ represents the anticlockwise rotation angle, and the minimum distance $g$ between the air cavity and line waveguide is maintained as 15 nm concurrently. For the sake of discussion, the edge index is defined as 1, 2, 3, 1', 2', 3', namely the existence of three directions along 11', 22', and 33'. Consequently, the $\theta$-dependent nUc spectrum under $\lambda$=1550 nm is depicted in Figure 3(b). nUc=8.08 at the peak of $P_2$ and nUc=8.09 at $P_3$ are greater than the case of $\theta$=0 (nUc=7.98). $P_2$ and $P_3$ denote the $\theta$ of

7.7 and 59.8 degrees, respectively. Figure 3(c) displays the λ-dependent nUc spectrum at P$_2$. The peak of nUc=8.10 emerges at point P$_4$ corresponding to λ=1549.9 nm. The $Q$ value of the peak is about 272.54. Additionally, the λ-dependent nUc spectrum at P$_3$ is demonstrated in Figure 3(d). The peak of nUc=8.62 occurs at point P$_5$ corresponding to λ=1549.3 nm. The $Q$ value of the peak is around 231.64.

Figure 3(e) illustrates the nE distribution at P$_4$. The maximum electric field is enhanced to 3.93. As a result, the fundamental WGM leads to the EM confinement inside the air cavity. The rotation has little impact on the WGM. More interestingly, the FP resonance emerges between the pair edges of 2 and 2'. On the other hand, the nE distribution at P$_5$ is shown in Figure 3(f). The maximum nE is 4.06. The WGM is inside the cavity, while the FP mode is in the direction of 11'. It should be noted that the FP mode is in the direction of 33' when θ is 52.8 degrees, which corresponds to the local maximum in Figure 3(b). According to the previous analysis, the evanescent electric field with different directions will enter the air cavity due to rotation, which changes the orientations of the FP mode. Finally, Figure 3(g) shows the λ-dependent R, A, and T spectra at P$_2$. At P$_4$, the maximum A is about 0.65, and R is around 0.15. According to the Fano lineshapes, the resonance wavelength is probably located around P$_4$, and the q for the T curve is about 0, for the A trend to infinity, and for the R close to -1, respectively. In addition, Figure 3(h) depicts the λ-dependent R, A, and T spectra at P$_3$. At P$_5$, the maximum A is about 0.65, and R is about 0.17. Therefore, at P$_4$ and P$_5$, the A values are greater and the R values are lower than those of non-rotated hexagonal and circular cavities.

**2.2 Hexagonal cavity-assisted exciting Mie resonances**

In this section, we utilize the rotated hexagonal cavity to assist in exciting the Mie resonances. The cavity is at P$_4$ with $l$=458 nm, θ=7.7°, and λ=1549.9 nm. An infinite-length silicon cylinder with a radius of $r_1$ is placed inside the center of the air cavity. Here, the refractive index of the lossless Si is regarded as a constant of 3.47 [56]. Firstly, we analyze the Mie scatterings of the isolated cylinder in free space, and Figure 4(a) is the schematic. The incident wavevector **k** (wavenumber $k$) is normal to the cylinder axis, the magnetic field (**H**) is parallel to the axis, and the wavelength is 1549.9 nm. Based on the Mie theory [57], the $i$th-order scattering efficiency can be written as

$$C_{sca} = \frac{2}{x}(|a_0|^2 + 2\sum_{i=1}^{\infty}|a_i|^2) \ . \tag{10}$$

Here, $a_i$ is the scattering coefficient read as

$$a_i = \frac{nJ'_i(x)J_i(nx) - J_i(x)J'_i(nx)}{nJ_i(nx)H'_i(x) - J'_i(nx)H_i(x)} \ , \tag{11}$$

where $J_i(x)$ and $H_i(x)$ correspond to the first kind Bessel and Hankel functions, $x=kr_1$, and $n$ denotes the refractive index of Si relative to the surroundings. Figure 4(b) depicts the total scattering efficiencies (tot, red curve) and the associated multipole contributions. $a_0$ to $a_3$ represent the magnetic dipole (MD, blue curve), ED (green curve), EQ (black curve), and electric octupole (EO, yellow curve) [48, 58], respectively. With increasing $r_1$, the MD, ED, EQ, and EO resonances emerge at the peaks of P$_6$ (166 nm), P$_7$ (253 nm), P$_8$ (344 nm), and P$_9$ (436 nm), respectively. The data marked by circles are calculated by FEM, which is consistent with the analytic analysis based on Mie theory. Figures 4(c)-4(f) demonstrate the scattering magnetic field |**H**$_z$| distributions of P$_6$ to P$_9$, respectively. The profile distributions of |**H**$_z$| validate the four Mie resonances.

Then, the Si cylinder (green circle) is placed into the cavity as shown in Figure 5(a). The nUc and $Q$ of air and Si are calculated based on the white and green domains in the cavity, respectively. When λ=1549.9 nm, Figure 5(b) illustrates the nUc spectra of air (red curve) and Si (blue curve) as a function of $r_1$. The red and blue curves possess four peaks simultaneously at points $P_{10}$-$P_{13}$. Correspondingly, $r_1$ is 175 nm, 303 nm, 310 nm, and 366 nm. The nUc values of air are greater than the Si at $P_{11}$ and $P_{13}$, while smaller at $P_{10}$ and $P_{12}$. The maximum nUc of air occurs at $P_{11}$, yet the case of Si is at $P_{12}$. Figures 5(c)-5(f) show the $|\mathbf{H}_z|$ distributions of $P_{10}$-$P_{13}$, respectively. Accordingly, we can verify the MD, EQ, ED, and EO resonances corresponding to $P_{10}$-$P_{13}$. Furthermore, Figures 5(c) and 5(f) indicate the obvious transmission at $P_{10}$ and $P_{13}$, meanwhile, the nUc values of these points are much lower than those of $P_{11}$ and $P_{12}$. Consequently, energy confinement is relatively small at $P_{10}$ and $P_{13}$. Therefore, we focus on the discussion at $P_{11}$ and $P_{12}$.

Figure 6(a) shows the $\lambda$-dependent nUc spectra at $P_{11}$. The maximum nUc of Si (blue curve) emerges at the peak of $P_{14}$ ($\lambda$=1527.2 nm). The $Q$ of the peak is 818.86. The maximum nUc of air (red curve) occurs at the peak of $P_{15}$ ($\lambda$=1548.5 nm). The $Q$ is about 461.28. Subsequently, Figure 6(b) demonstrates the $\lambda$-dependent R, A, and T at $P_{11}$. At $P_{14}$, maximum A is 0.63, and R is about 0.15. At $P_{15}$, maximum A is about 0.65, and R is about 0.15. Therefore, the strong EM confinement in air or Si domains always generates maximum absorption and relatively low reflection. The subtle difference is that A is somewhat higher at $P_{15}$ than at $P_{14}$. Likewise, the R, A, and T curves exhibit typical Fano lineshapes, and the resonant wavelength locations and q values analysis are similar to Figure 3(g). Differently, Mie resonances are involved in the mode hybrid. Figures 6(c) and 6(d) depict the $|\mathbf{H}_z|$ distributions of $P_{14}$ and $P_{15}$, respectively. As a result, ED resonance at $P_{14}$ results in maximum nUc in the Si cylinder, whereas the EQ resonance at $P_{15}$ generates maximum nUc in the remaining air. Figures 6(e) and 6(f) show the nE distributions of $P_{14}$ and $P_{15}$, respectively. The FP resonance at $P_{14}$ is in the direction of 22', while the dominant FP resonance at $P_{15}$ is in the direction of 33'. Therefore, different Mie resonances will change the localized directions of FP modes even in the same cavity.

Similarly, the nUc spectra concerning $\lambda$ at $P_{12}$ are depicted in Figure 7(a). The maximum nUc of Si is obtained at $P_{16}$ ($\lambda$=1549.3 nm), and the $Q$ is about 752.78. The maximum nUc of air is achieved at $P_{17}$ ($\lambda$=1591.3 nm) with $Q$ being about 500.28. Subsequently, Figure 7(b) demonstrates the $\lambda$-dependent R, A, and T. At $P_{16}$, the maximum A is about 0.64, while R is around 0.14. At $P_{17}$, the maximum A is about 0.67, and R is about 0.11. The $|\mathbf{H}_z|$ distributions of $P_{16}$ and $P_{17}$ are shown in Figures 7(c) and 7(d), respectively. Therefore, the maximum nUc of Si originates from the ED resonance, whereas the air one is due to the EQ resonance. Figures 7(e) and 7(f) display the nE distributions of $P_{16}$ and $P_{17}$, respectively. At $P_{16}$, the FP resonance is oriented in the direction of 22', while it emerges in the direction of 33' at $P_{17}$.

Repeat the preceding steps based on the identical cavity at $P_4$, yet $r_1$ varies with an interval of 1 nm. Figure 8(a) shows the maximum nUc spectra of Si (blue triangle) and air (red circle). Both the curves possess the minimum values at $r_1$=295 nm. Furthermore, the curves exhibit the Fano-like lineshape with q values tending to 0 at the dips. Figure 8(b) indicates the wavelength responses of the maximum nUc of Si (blue triangle) and air (red circle). The blue line, with a slope of 3.24, fits the Si data, and the red line, with a slope of 5.74, fits the air data. Both wavelengths show linear responses, while the air one changes faster. The cross point occurs near $r_1$=295 nm. The linear relationship between the wavelength and the size is consistent with the scaling properties of the Maxwell equations [46]. Moreover, based on the preceding analysis, the wavelength of maximum nUc of Si corresponds to ED resonance, while the air one corresponds to EQ resonance. Therefore, the ED and EQ resonances overlap at the cross point, resulting in the hybrid modes of ED, EQ, FP, WGM, and SPPs.

Then, Figure 8(c) depicts the $Q$ spectra of the maximum nUc of Si and air. The curves exhibit Fano-like lineshapes, with a positive q for the Si case, but a negative q for the air case. The maximum $Q$ of Si is about 1160.4 at $r_1$=296 nm, and that of air is about 810.1 at $r_1$=294 nm. However, both $Q$ values are relatively low at $r_1$=295 nm. Next, Figure 8(d) shows the R spectra at the maximum nUc. The Fano-like curves exhibit a negative q for the Si case, but a positive q for the air case. Thus, the minimum R of Si (about 0.096) occurs at $r_1$=298 nm, yet that of air (about 0.054) appears at $r_1$=294 nm. Then, the Fano-like T curves are demonstrated in Figure 8(e). q for the Si is positive, but is negative for the air. The dramatic change emerges around $r_1$=295 nm. Finally, Figure 8(f) shows the Fano-like A curves of the two cases. Both q values are close to 0, and the dips occur at $r_1$=296 nm for the Si and at $r_1$=294 nm for the air. However, on the right sides of the dips, the larger $r_1$ achieves larger absorption.

**2.3 Rotated Hexagonal cavity-assisted THG**

Based on the linear responses of the Si cylinder inside the rotated hexagonal cavity, we will explore the nonlinear responses of the same system. The schematic diagram is shown in Figure 9. The cavity has the same conditions of $P_4$, the red arrow indicates the incident power ($P_{FWin}$) of the fundamental wave (FW), the blue arrow denotes the transmitted power ($P_{THGT}$) of the THG, and the green arrow represents the reflected power ($P_{THGR}$) of the THG. Accordingly, the reflected ($\eta_R$) and transmission ($\eta_T$) THG conversion efficiencies are expressed by

$$\eta_R = \frac{P_{THGR}}{P_{FWin}} \ , \ \eta_T = \frac{P_{THGT}}{P_{FWin}} \ . \tag{12}$$

All powers are calculated by integrating the energy density along the input and output ports. Here, we focus on the THG of the centrosymmetric Si cylinder because the THG of the Ag plate is very weak [27]. The third-order nonlinearity is induced by the nonlinear polarization vector $\mathbf{p}^{(3)}$ written as [59]

$$\mathbf{p}^{(3)} = \varepsilon_0 \chi^{(3)} (\mathbf{E}(\omega) \cdot \mathbf{E}(\omega)) \mathbf{E}(\omega) \ , \tag{13}$$

where the third-order nonlinear susceptibility $\chi^{(3)}$ of Si is about $3.49 \times 10^{-18}$ m$^2$/V$^2$ [60]. Theoretically, the electric field of the THG is deduced by substituting the $\mathbf{p}^{(3)}$ into the Maxwell Equations, then the THG power can be calculated accordingly. The incident intensity of the fundamental wave is set to be 55 MW/cm$^2$.

In the first case, the refractive index of the Si (green region) is still 3.47. The linear responses are identical to the analysis of section 2.2. Then Figure 10(a) shows the $\eta_R$ spectrum as a function of $r_1$. The two peaks appear at $r_1$=303 nm ($P_{11}$) and $r_1$=310 nm ($P_{12}$). The $\eta_R$ of $P_{12}$ is about 6 times that of $P_{11}$, originating from the intensity inside the cylinder of $P_{12}$ being about 3.37 times that of $P_{11}$, as shown in Figure 5(b). Next, the $r_1$-dependent $\eta_T$ spectrum is demonstrated in Figure 10(b). Similarly, the $\eta_T$ of $P_{12}$ is about 56.48 times that of $P_{11}$. In comparison, the $\eta_R$ of $P_{12}$ is almost the same as its $\eta_T$, while the $\eta_R$ of $P_{11}$ is much greater than its $\eta_T$. Therefore, the ED resonance (at $P_{12}$) confines more energy inside the Si cylinder than the EQ resonance (at $P_{11}$), resulting in much stronger THG at $P_{12}$.

Subsequently, Figures 10(c) and 10(d) depict the $\eta_R$ and $\eta_T$ spectra at $P_{11}$ as a function of the fundamental wavelength, respectively. The top horizontal coordinate axis displays the wavelength of the THG. As a result, the pronounced $\eta_R$ and $\eta_T$ peaks occur around 1527 nm, yet quite weak efficiencies appear around 1549 nm. The maximum efficiency is $\eta_T$=9.57×10$^{-5}$, and the corresponding THG wavelength is 509.12 nm. Figure 6(a) shows that the nUc is much larger in the cylinder at around 1527 nm, but much larger in the air at around 1549 nm. According to Figures 6(c) and 6(d), the ED resonance of the shorter wavelength is more advantageous to THG than the EQ resonance of the longer one. Figures 10(g) and 10(h) show the $|\mathbf{H}_z|$ and nE distributions of the third harmonic wave at the peak, respectively. Therefore, the THG can be further confirmed through the nodal lines and points in the radial and azimuthal directions of the distributions.

Analogously, Figures 10(e) and 10(f) show $\eta_R$ and $\eta_T$ spectra at $P_{12}$. The peaks emerge around 1549 nm, while very weak responses appear around 1592 nm. The maximum efficiency is $\eta_R \sim \eta_T \sim 3.32 \times 10^{-5}$, and the corresponding THG wavelength is 516.48 nm. Figure 7(a) shows that the Si cylinder has the maximum nUc at around 1549 nm. According to Figures 7(c) and 7(d), the ED resonance at the short wavelength aids in the THG. The $|\mathbf{H}_z|$ and nE distributions of the third harmonic wave at the peak are demonstrated in Figures 10(i) and 10(j), respectively. Consequently, the THG is further verified.

Interestingly, what happens to the THG under $r_1$=295 nm? Figure 8 indicates the overlap of ED and EQ resonances at this condition, leading to the minimum energy confinement in the cylinder. Figures 11(a) and 11(b) illustrate $\eta_R$ and $\eta_T$ curves of this case concerning the fundamental wavelength, respectively. Thus, the maximum THG efficiency is $\eta_T \sim 2.74 \times 10^{-6}$, only 1/35 times the preceding maximum THG efficiency. Then, the curves exhibit the typical Fano lineshapes, and the minimum dips emerge at 1501.3 nm, i.e., the wavelength of the cross point in Figure 8(b). Consequently, the overlap of ED and EQ resonances decreases the THG efficiency.

Finally, we study the impact of incident intensity ($I_{in}$) on THG efficiency. We focus on the case of $P_{11}$. When $I_{in}$ ranges from 30 to 80 MW/cm$^2$ with a step of 5 MW/cm$^2$, Figures 12(a) and 12(b) show the $\eta_R$ and $\eta_T$ spectra as a function of the incident wavelength, respectively. The peak location does not shift, while the peak efficiency increases as the intensity increases. Figure 12(c) demonstrates the peak efficiency spectra of $\eta_R$ (red square) and $\eta_T$ (blue circle) concerning $I_{in}$. The red and blue lines are their fitting results, respectively. The fitting power function is $\eta_{R(T)} = a(I_{in})^b$. Here, $a$=2.4×10$^{-8}$ and $b$=2 for the red line, and $a$=3.165×10$^{-8}$ and $b$=2 for the blue one. Based on the formula 12 and the $b$=2, the power of THG is the cubic order of the incident fundamental wave power, validating the THG again [61].

The refractive index of Si has slightly larger real and imaginary parts at the green light than in the near-infrared [56]. In the second case, we consider the THG from a practical Si cylinder. The cavity agrees with the conditions of $P_4$ with the operating wavelength of 1549.9 nm. Figure 13(a) shows the nUc spectra of Si (blue line) and air (red line) as a function of $r_1$. With $r_1$ increasing, the four peaks correspond to the MD, EQ, ED, and EO resonances, consistent with the prior analysis in Figure 5. Differently, the maximum nUc of air emerges at $r_1$=366 nm due to the EO resonance, and the maximum nUc of Si occurs at $r_1$=308 nm (smaller than the $P_{12}$). Then, Figures 13(b) and 13(c) display the $\eta_R$ and $\eta_T$ spectra, respectively. At $r_1$=308 nm, the two curves have their peaks simultaneously, while the peak value of $\eta_T$ is somewhat higher. Figure 13(d) illustrates the $|\mathbf{H}_z|$ distribution of the fundamental wave. As a result, ED resonance is excited at this radius. The $|\mathbf{H}_z|$ distribution of the third-order harmonic wave in Figure 13(e) verifies the THG. Figures 13(f) and 13(g) show the $\eta_R$ and $\eta_T$ spectra at that radius, respectively. The maximum $\eta_R$ and $\eta_T$ occur at 1549.2 nm, and the THG wavelength is 516.4 nm. Then the maximum efficiency is $\eta_T \sim 3.91 \times 10^{-7}$. The maximum efficiency decreases by two orders of magnitude compared to the constant case of $P_{12}$.

## 2.4 Influence of cavity thickness

Finally, we consider a manufacturable 3D rotated cavity under the conditions of $P_4$. Figure 14(a) is the schematic, and the thickness d is $c_d \times \lambda$. Here, $c_d$ is a thickness coefficient. Figure 14(b) demonstrates the nUc spectra as a function of $c_d$. As a result, the ultra-thin structure cannot efficiently confine energy, while the confinement effect becomes significant as the thickness increases. The reason is that the ultra-thin structure cannot support the WGM and FP resonance. In addition, Figure 14(c) illustrates the R, A, and T spectra concerning $c_d$. With $c_d$ increasing, the R decreases slowly, A increases gradually, and T increases dramatically first and then becomes a modest slowdown. Therefore, when the thickness far exceeds the incident wavelength, the response is closer to the ideal case obtained in the previous chapters.

**Conclusion**

In summary, we have proposed a rotated hexagonal air cavity integrated within an Ag-air-Ag waveguide operating at the near-infrared regime and realized the THG of green light assisted by the hybrid modes. Compared with the circular cavity, the hexagonal cavity introduces FP resonance, leading to much stronger energy confinement with a large-scale 'hot area' of evanescent wave. The orientation of the FP resonance can be tuned by rotating the cavity. The hybrid modes of FP resonance, SPPs, and WGM result in the filtering with maximum absorption and quite low reflection. Then, putting a Si cylinder into the cavity, the confined evanescent electric field can efficiently excite resonant Mie multipoles. As a result, the ED resonance leads to significant energy confinement inside the cylinder, whereas the EQ resonance leads to confinement in the remaining air. However, the overlap of ED and EQ resonances diminishes the confinement. More interestingly, the hybrid modes with Mie resonances produce typical Fano-like lineshapes. Subsequently, remarkable cavity-assisted THG occurs when ED resonance is excited. Then, the efficiency of THG is the square of the incident intensity, and the power of THG is the cube of the incident power. Finally, THG efficiency decreases by two orders when adopting the actual Si material due to the absorption.

The proposed cavity has two advantages: large-scale incident electric field enhancement and less backward reflection filtering. Unlike the THG excited directly in free space [60, 62-65], our cavity with a hot area can enhance the light-matter interaction, which is expected to reduce the excitation intensity threshold for nonlinear processes. The suppression of backward reflection is favorable to chip optical communication. Furthermore, our fillable cavity provides a flexible auxiliary component to generate other nonlinear optical responses, such as filling active materials to observe the fluorescence and Raman effects, or filling AlGaAs particles to excite the second harmonic wave [66, 67]. Moreover, our cavity offers a simple way of adjusting the radius of the Si cylinder to generate continuously adjustable THG because of the linear relationship between the pumping wavelength and the radius. However, to further improve the efficiency of the nonlinear responses, we need to trade off the field enhancement and the energy dissipation. Next, we anticipate further boosting confinement and efficiency by implementing multi-cavity configurations [68] or adopting an inverse algorithm based on our design principle [10, 69], while using polygon modes induced by shape perturbation to reduce the dissipation [11]. Overall, our work paves the way to design a high-performance air cavity, provides a versatile platform for generating linear and nonlinear optical phenomena, and has potential applications in optical communication, sensors, and miniature spectrometers [69, 70] and lasers [71].

**Acknowledgments**

This work was supported by the National Natural Science Foundation of China (12304425, 12434017, 12374355); Sichuan Science and Technology Program (2024NSFSC1354).

**Disclosures**

The authors declare no competing financial interest.

**Data availability.**

Data may be obtained from the authors upon reasonable request.

**Figures**

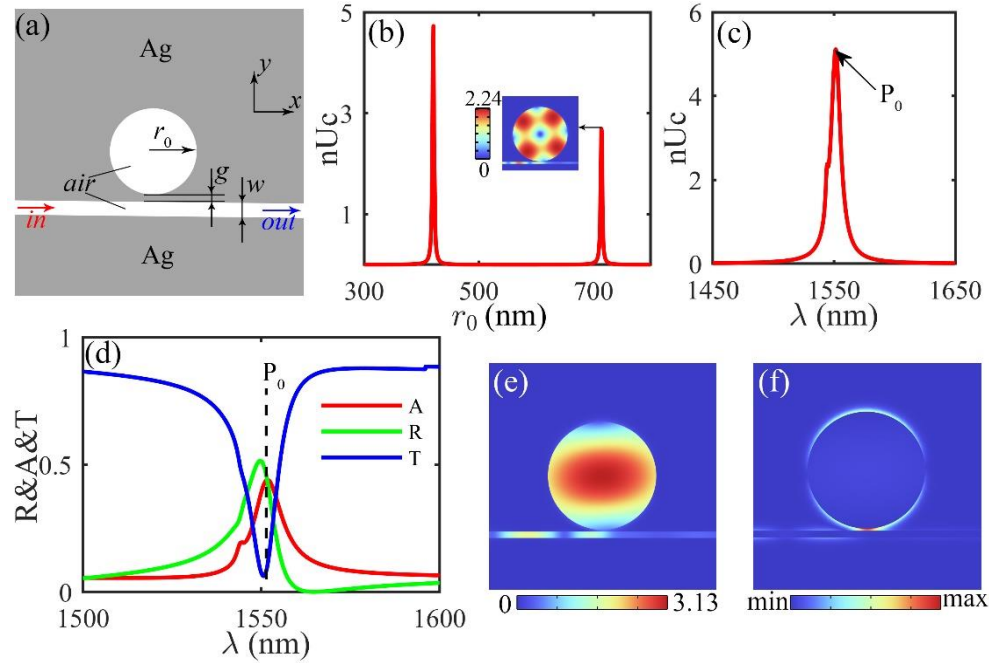

Figure 1. A circular air cavity is integrated into the Ag-air-Ag waveguide. (a) Schematic of the structure. The red and blue arrows correspond to the energy input and output. (b) When λ=1550 nm, nUc as a function of $r_0$. The inset is the electric field norm nE (V/m) distribution at the second peak. (c) λ-dependent nUc when $r_0$=421 nm (the first peak in (b)). $P_0$ denotes the peak wavelength of λ=1551.4 nm. (d) Reflectivity (R), absorption (A), and transmissivity (T) spectra concerning λ. At $P_0$ (i.e., $r_0$=421 nm, and λ=1551.4 nm fixed), (e) the nE distribution, and (f) the current density norm nJ distribution, the unit is A/m$^2$.

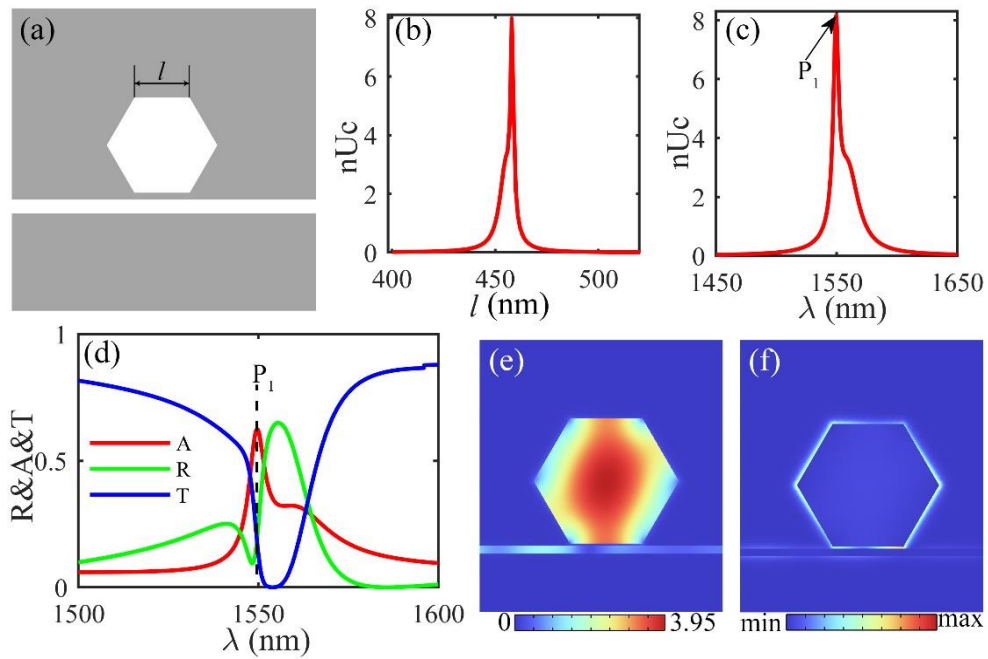

Figure 2. A hexagonal air cavity is integrated into the Ag-air-Ag waveguide. (a) Schematic of the structure. $l$ is the side length. (b) When λ=1550 nm, nUc as a function of $l$. (c) λ-dependent nUc at the peak in (b). Point $P_1$ denotes the peak wavelength of λ=1549.6 nm. (d) R, A, and T spectra concerning λ when $r_0$=458 nm. (e) and (f) the nE and nJ at point $P_1$, respectively.

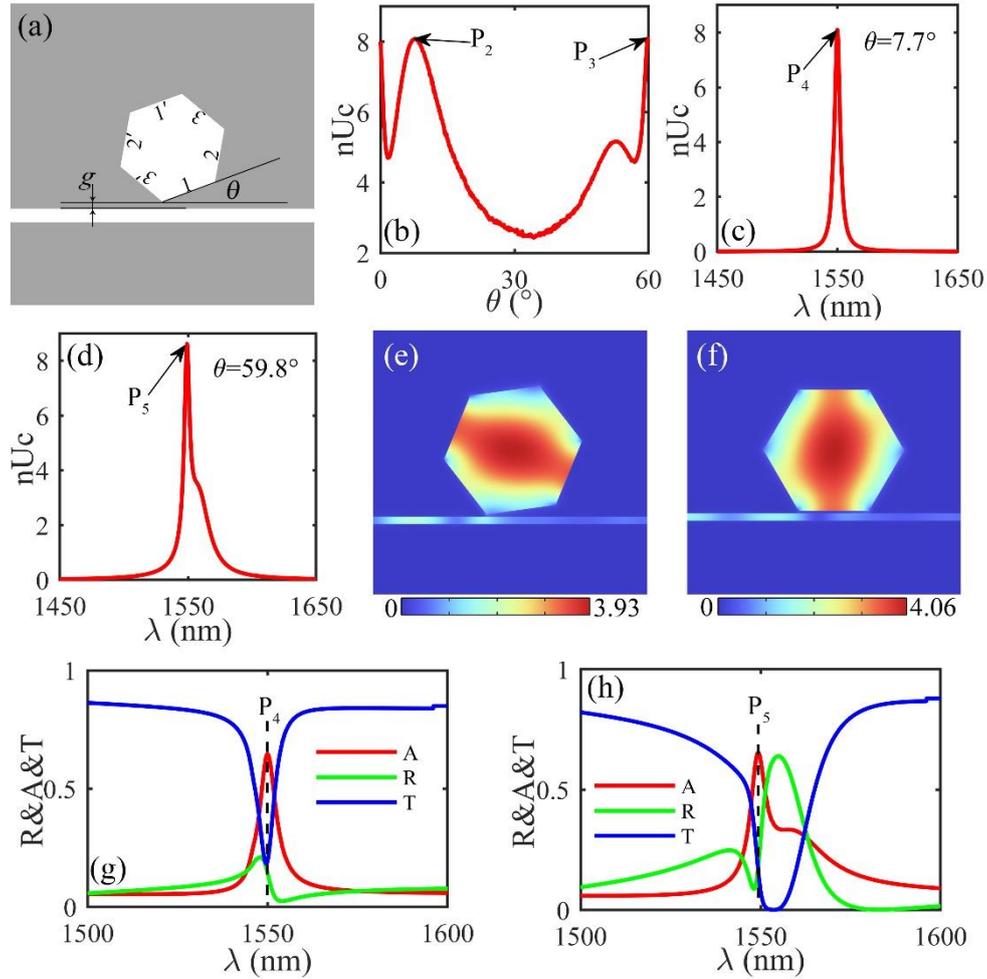

Figure 3. A rotated hexagonal air cavity is integrated into the Ag-air-Ag waveguide. (a) Schematic of the structure. $\theta$ is the rotation angle. 1, 2, 3, 1', 2', 3' denote the indices of the sides. (b) nUc as a function of $\theta$ when $r_0$=458 nm and $\lambda$=1550 nm. Point $P_2$ denotes $\theta$=7.7° and $P_3$ denotes $\theta$=59.8°. (c) $\lambda$-dependent nUc at point $P_2$. Point $P_4$ denotes $\lambda$=1549.9 nm. (d) $\lambda$-dependent nUc at point $P_3$. Point $P_5$ denotes $\lambda$=1549.3 nm. (e) and (f) the nE distributions at $P_4$ and $P_5$, respectively. (g) and (h) $\lambda$-dependent R, A, and T at $P_2$ and $P_3$, respectively.

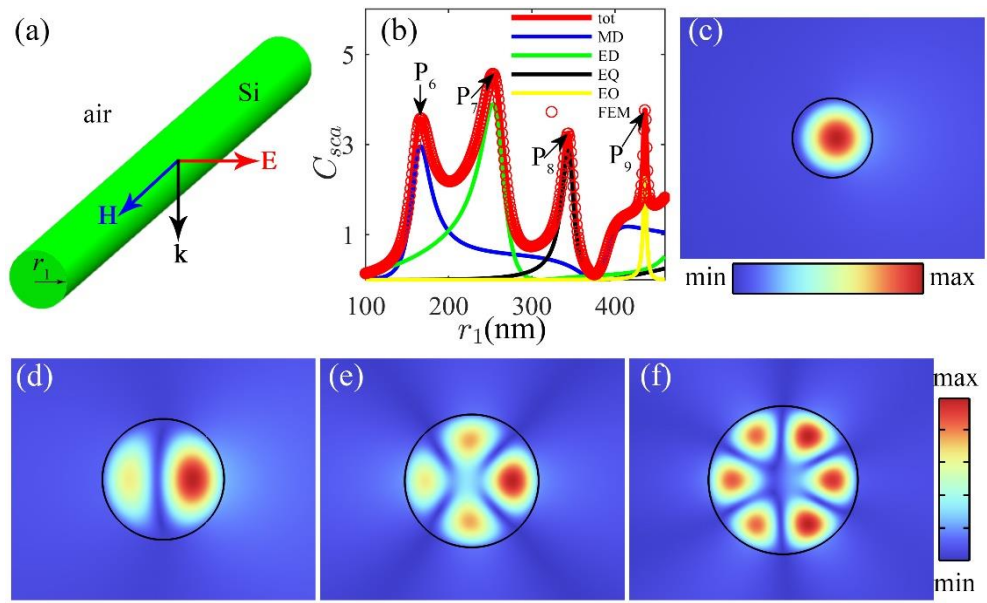

Figure 4. Scattering of an isolated Si cylinder in air. (a) The schematic, radius $r_1$, normally incident wave with wavevector (**k**), magnetic field (**H**) parallel to the center axis of the cylinder. (b) Scattering efficiencies ($C_{sca}$) and the multipole contributions as a function of $r_1$. (c)-(f) Scattering magnetic field |$H_z$| distributions of points $P_6$-$P_9$, respectively.

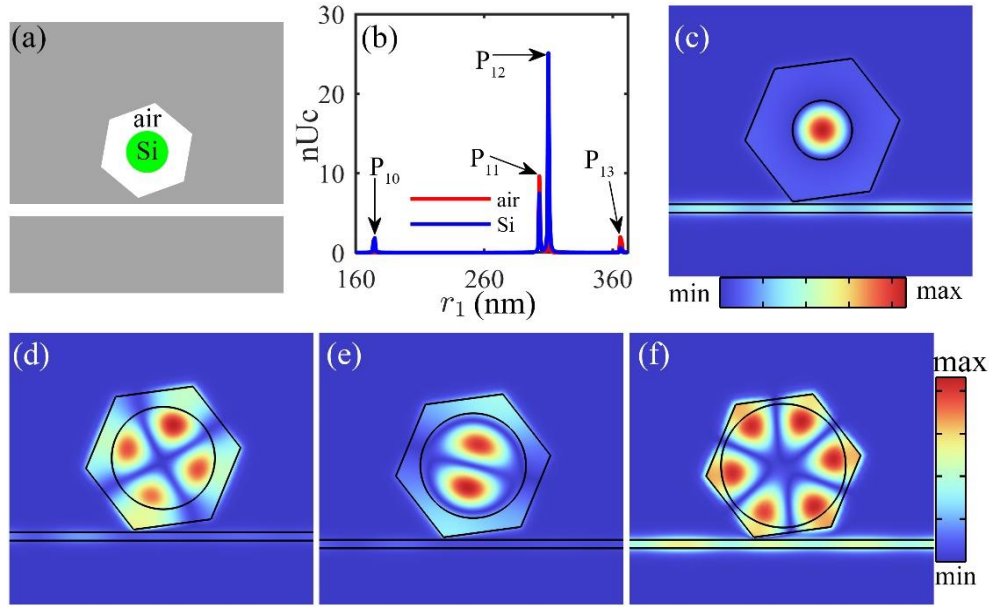

Figure 5. Putting a Si cylinder into the rotated hexagonal air cavity under the conditions of $P_4$. (a) The schematic. The green domain is Si, and the white is air. (b) nUc of the cylinder (Si, blue curve) and remaining air (air, red curve) as a function of $r_1$. (c)-(f) |$H_z$| distributions of $P_{10}$-$P_{13}$, respectively.

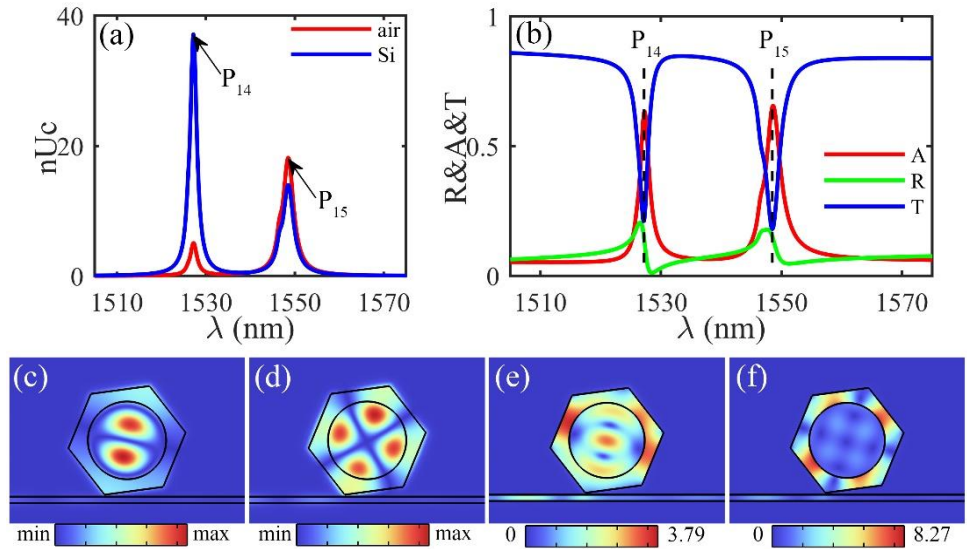

Figure 6. Responses of the Si cylinder filled rotated hexagonal cavity at $P_{11}$. (a) nUc as a function of λ. (b) R, A, and T spectra as a function of λ. (c)-(d) |$H_z$| distributions of $P_{14}$ and $P_{15}$, respectively. (e)-(f) nE distributions of $P_{14}$ and $P_{15}$, respectively.

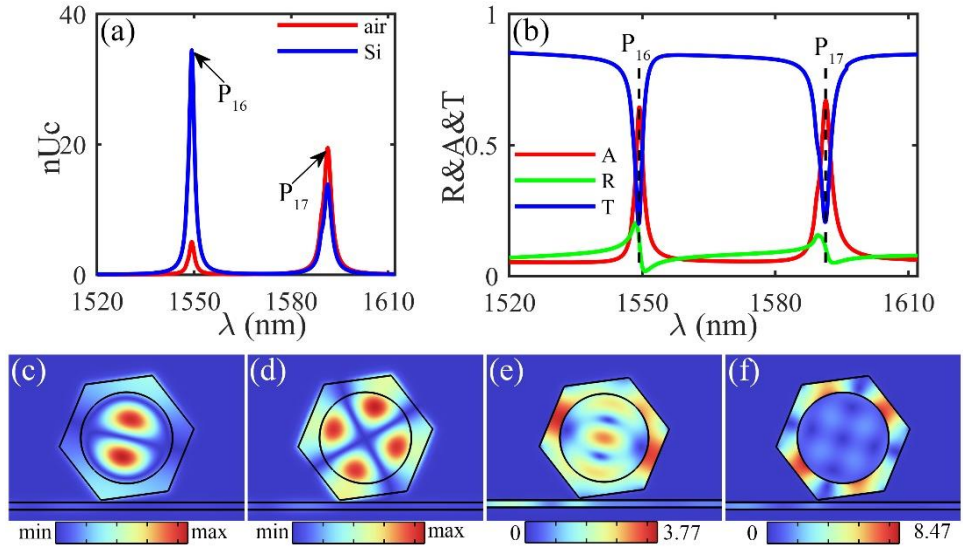

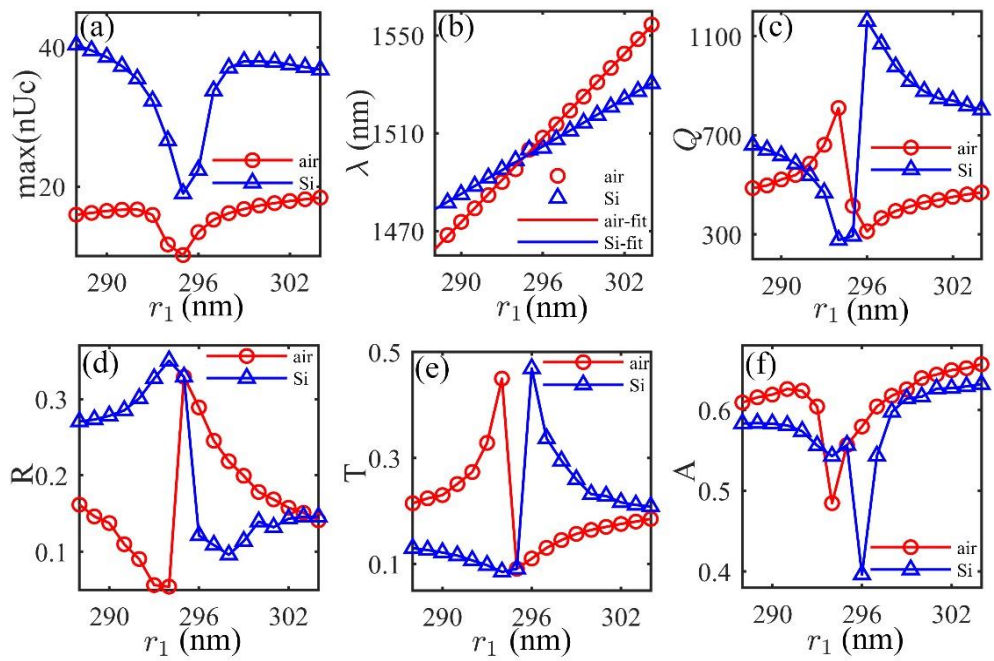

Figure 7. Responses of the Si cylinder filled rotated hexagonal cavity at $P_{12}$. (a) nUc as a function of $\lambda$. (b) R, A, and T spectra as a function of $\lambda$. (c)-(d) $|H_z|$ distributions of $P_{16}$ and $P_{17}$, respectively. (e)-(f) nE distributions of $P_{16}$ and $P_{17}$, respectively.

Figure 8. Responses of the Si cylinder filled rotated hexagonal cavity with $r_1$ change. (a) max(nUc) of Si and the remaining air. (b)-(f) The spectra of $\lambda$, $Q$, R, T, A corresponding to the max(nUc), respectively.

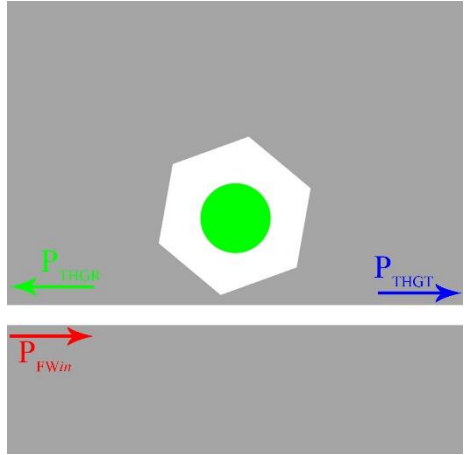

Figure 9. Schematic of the Si cylinder filled rotated hexagonal cavity under the conditions of $P_4$. $P_{FWin}$ represents the power of the incident fundamental wave (FW). $P_{THGR}$ and $P_{THGT}$ are the THG powers at the input and output ports, respectively.

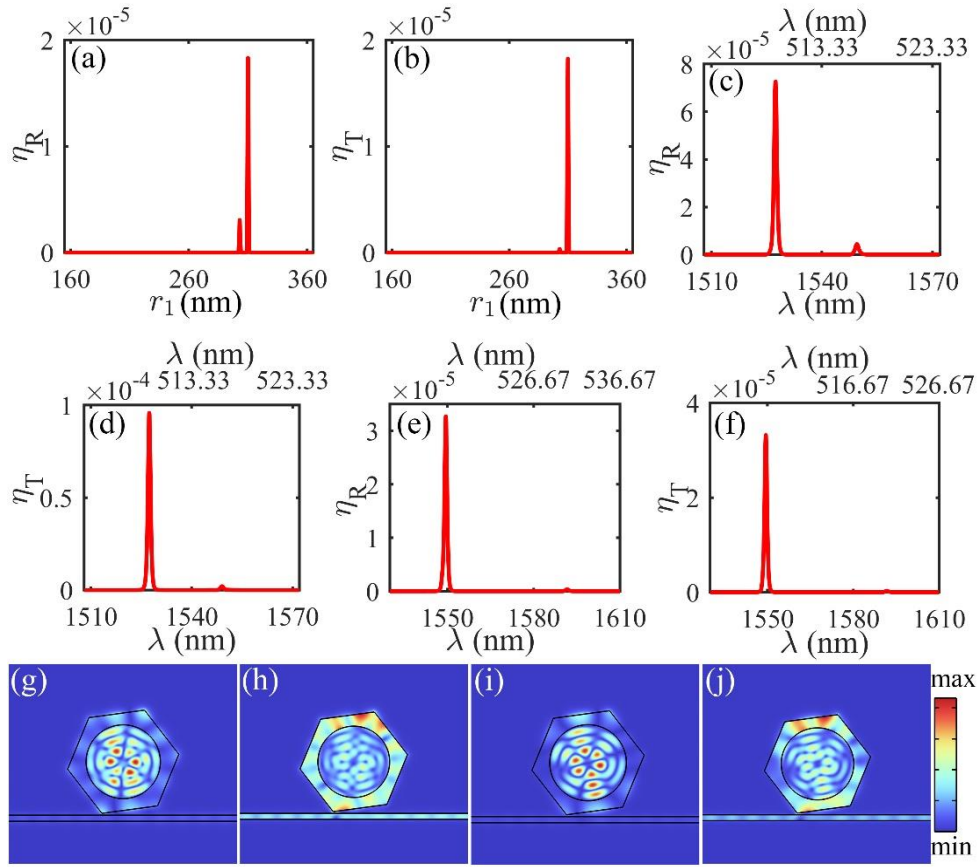

Figure 10. The THG of the Si cylinder filled rotated hexagonal cavity under the conditions of $P_4$. (a) and (b) The THG efficiencies of the input ($\eta_R$) and output ($\eta_T$) ports as a function of $r_1$, respectively. The top horizontal coordinate axis displays the harmonic wavelength. (c) and (d) The $\eta_R$ and $\eta_T$ at $P_{11}$ ($r_1$=303 nm). (e) and (f) The $\eta_R$ and $\eta_T$ at $P_{12}$ ($r_1$=310 nm). (g) and (h) |$H_z$| and nE distributions of the THG at the peak in (d). (i) and (j) |$H_z$| and nE distributions of the THG at the peak in (f).

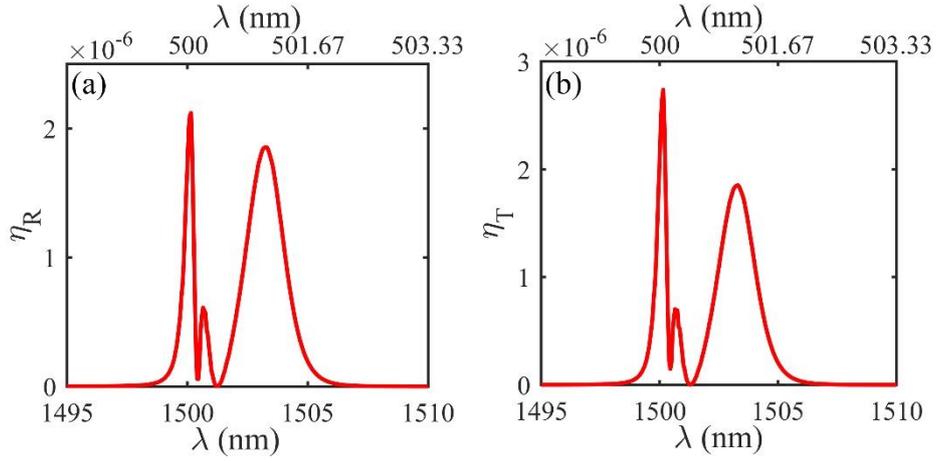

Figure 11. (a) and (b) The $\eta_R$ and $\eta_T$ of the Si cylinder filled rotated hexagonal cavity with $r_1$=295 nm.

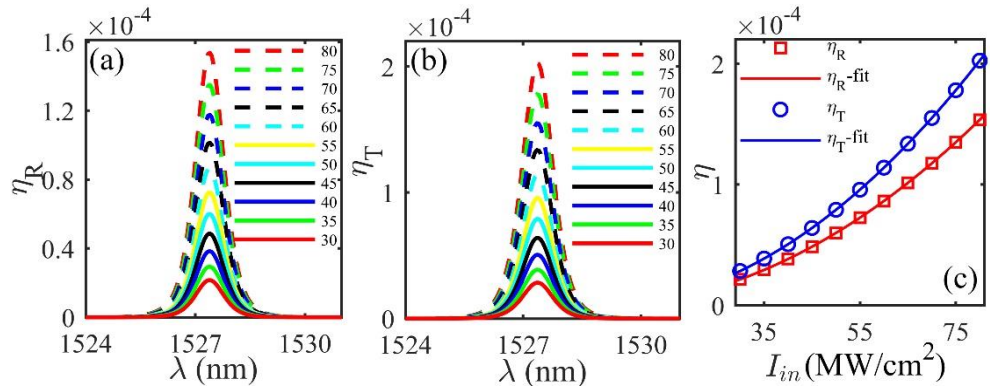

Figure 12. The THG efficiency as a function of incident intensity ($I_{in}$). (a) and (b) The $\eta_R$ and $\eta_T$ concerning λ at different $I_{in}$. (c) The peak $\eta_R$ and $\eta_T$ in (a) and (b) as a function of $I_{in}$, respectively. $\eta_R$-fit and $\eta_T$-fit are the fitting lines.

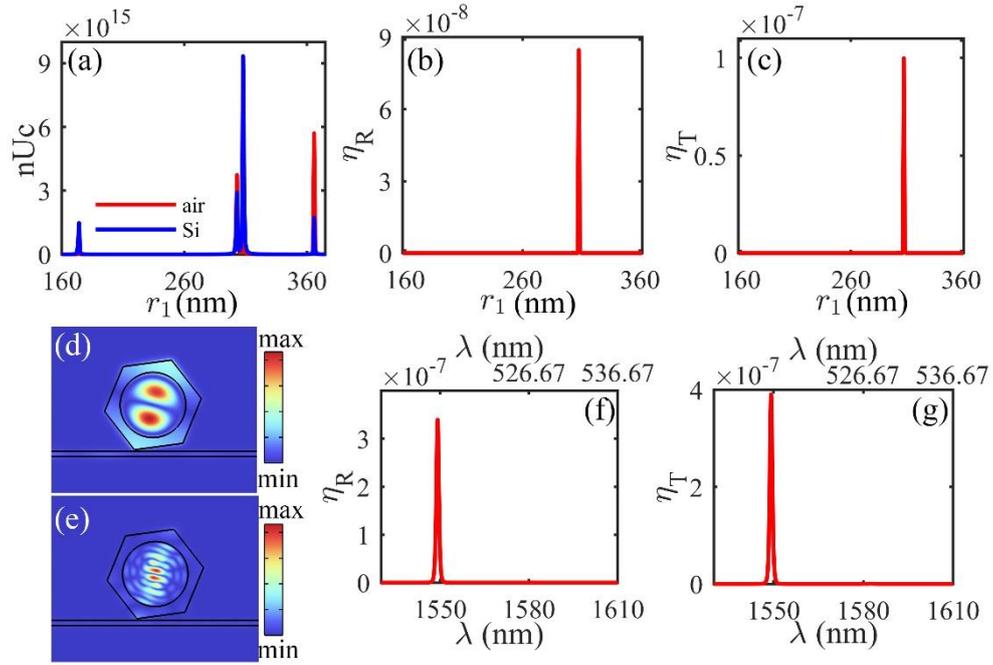

Figure 13. The THG of the Si cylinder filled rotated hexagonal realistic cavity under the conditions of $P_4$. When $\lambda=1549.9$ nm, (a) nUc, (b) $\eta_R$, and (c) $\eta_T$ as a function of $r_1$, respectively. (d) and (e) $|\mathbf{H}_z|$ distributions of the fundamental and harmonic waves at the peak in (c). (f) and (g) $\lambda$-dependant $\eta_R$ and $\eta_T$ when $r_1=308$ nm, respectively.

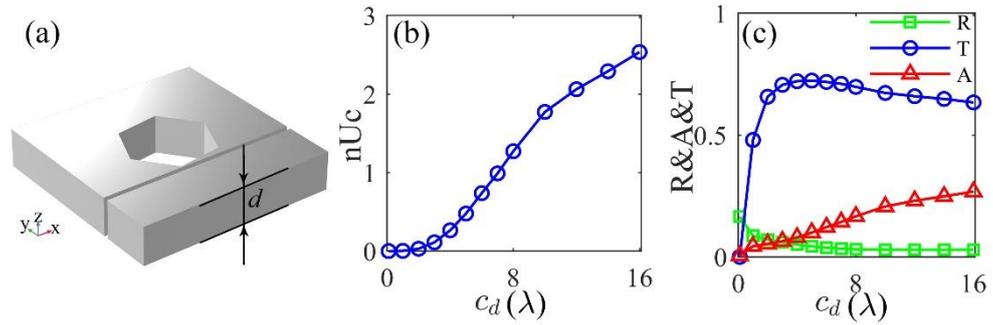

Figure 14. The thickness effect. (a) Schematic of the rotated hexagonal air cavity at $P_4$ with a thickness of $d$. (b) nUc as a function of a thick coefficient $c_d$. $d=c_d*\lambda$. (c) R, A, T spectra as a function of $c_d$.